\documentclass[conference]{IEEEtran}

\ifCLASSINFOpdf
   \usepackage[pdftex]{graphicx}
   \graphicspath{{../pdf/}{../jpeg/}}
   \DeclareGraphicsExtensions{.pdf,.jpeg,.png}
\else
   \usepackage[dvips]{graphicx}
   \graphicspath{{../eps/}}
   \DeclareGraphicsExtensions{.eps}
\fi

\usepackage{amsmath}
\usepackage{bm}

\usepackage{subfigure}

\usepackage{amssymb}
\usepackage{makecell}
\usepackage{color}
\usepackage{multirow}
\usepackage{arydshln}
\usepackage{amsthm}

\theoremstyle{remark}
\newtheorem{example}{Example}

\begin{document}

\title{A Novel Iterative Soft-Decision Decoding Algorithm for RS-SPC Product Codes}

\author{Mingyang~Zhu$^{1}$,
        Ming~Jiang$^{1, 2}$,
        Chunming~Zhao$^{1, 2}$\\
        $^{1}$National Mobile Communications Research Lab., Southeast University, Nanjing 210096, China\\
        $^{2}$The Purple Mountain Laboratories, Nanjing, China\\
        Email:~\{zhumingyang, jiang\_ming, cmzhao\}@seu.edu.cn\thanks{}}
\maketitle

\begin{abstract}
This paper presents a generalized construction of RS-SPC product codes. A low-complexity joint-decoding scheme is proposed for these codes, in which a BP-based iterative decoding is performed based on the binary expansion of the whole parity-check matrix. Various powerful RS codes can be used as the component codes for RS-SPC product codes, which gives a good performance for local decoding (decode a single component codeword). The proposed BP-based iterative decoding is a global decoding, and it achieves an error-correcting capability comparable to codes of large blocklengths. This two-phase decoding scheme preserves the low decoding latency and complexity of the local decoding while achieves high reliability through the global decoding. The complexity of the proposed iterative decoding is discussed, and the simulation results show the proposed scheme offers a good trade-off between the complexity and the error performance.
\end{abstract}

%\begin{IEEEkeywords}
%\textit{q}-ary spatially coupled LDPC codes, iterative decoding threshold, global and local decoding.
%\end{IEEEkeywords}
\IEEEpeerreviewmaketitle

\section{Introduction}
Reed-Solomon (RS) codes \cite{Reed_Solomon}--\cite{LinShu} are one of the most important \textit{maximum distance separable} codes and they are widely used in many communication and storage systems. In most of these existing systems, RS codes are decoded by algebraic hard-decision decoding (HDD) algorithms, such as the Berlekamp-Massey (denoted by BM-HDD) algorithm \cite{Berlekamp}. However, HDD does not utilize soft information thus it usually incurs a significant performance loss compared to a suitable soft-decision decoding (SDD) algorithm. In order to make use of the soft information, the generalized minimum distance (GMD) decoding \cite{GMD}, Chase decoding \cite{Chase}, and algebraic soft-decision (ASD) decoding \cite{KV} algorithms were proposed. These SDD algorithms improve performance over the traditional HDD method, however, the performance gaps between them and the \textit{maximum-likelihood decoding} (MLD) are still noticeable particularly for long RS codes.

Iterative decoding algorithms based on belief propagation (BP), e.g., the sum-product algorithm (SPA) and the min-sum algorithm (MSA), are widely used for decoding the low-density parity-check (LDPC) codes. However, the parity-check matrices of RS codes are in general not sparse, thus directly applying the iterative decoding algorithms to RS codes is difficult. In order to deal with such high-density parity-check (HDPC) matrices, Jiang and Narayanan proposed a BP algorithm by adapting the parity-check matrix (denoted by ABP) \cite{JN_ABP}, where Gaussian elimination is applied before each iteration, thus its complexity may not be tolerated in practical applications. In \cite{SSID}, a BP algorithm by stochastically and cyclically shifting (denoted by SSID) was proposed to enhance the conventional BP algorithm for RS codes. This algorithm is easy to realize, whereas, it performs not well for the RS codes over large fields, e.g., GF($2^p$) with $p > 6$.

Although there is no particularly effective SDD scheme for a single RS codeword, the SDD of an RS-based concatenated coding scheme improves the error performance a lot even over the MLD of a single RS codeword, for example, the turbo product codes (TPCs) \cite{TPC_Pyndiah} with RS component codes and the cascaded RS codes \cite{GFT_ISDD} which are jointly encoded through \textit{Galois Fourier transform}. However, the decoding can not perform for a single component codeword for the cascaded RS codes in \cite{GFT_ISDD} due to the interleaving and the Galois Fourier transform. The decoding of a single component codeword and the whole codeword are denoted as the local decoding (LD) and the global decoding (GD), respectively. This two-phase decoding scheme preserves the low decoding latency of the LD while the error-correcting capability of the GD is comparable to codes of large blocklengths.

In this paper, we present a low-complexity scheme for jointly encoding and decoding RS codes. The encoding is a direct product of RS codes and single-parity-check (SPC) codes thus a code of this type is called an RS-SPC product code. The conventional decoding scheme of product codes is a turbo process of the row and column decoders. We develop a BP-based iterative decoding scheme for RS-SPC product codes based on the binary expansion of the parity-check matrix. An important advantage of this code is the error performance of the LD in general improves over the LD of TPCs. TPCs usually consist of 1 or 2-error-correcting component codes, while more powerful component codes are used for RS-SPC product codes. Moreover, the presented BP-based iterative decoding scheme achieves an error performance comparable to long-blocklength codes. The iterative decoding can be highly parallelized with a relatively low complexity. The simulation results show the RS-SPC product codes offer a good trade-off between the LD (i.e., complexity) and the GD (i.e., performance).

\section{RS-SPC Product Code}
Consider a narrow sense $t$-error-correcting $(n,k,\delta)$ RS code over GF($2^p$) with $n = 2^p - 1$ and a minimum distance $\delta=n-k+1=2t+1$. The parity check matrix over GF($2^p$) is given by the $(n-k) \times n$ matrix:
\begin{equation}
\label{RS_Standard_H}
{{\bf{H}}_s} = \left[ {\begin{array}{*{4}{c}}
1&\beta&\cdots&{\beta}^{n-1}\\
1&{\beta}^2&\cdots&{\beta}^{2(n-1)}\\
\vdots&\vdots&\ddots&\vdots\\
1&{\beta}^{n-k}&\cdots&{\beta}^{(n-k)(n-1)}
\end{array}} \right],
\end{equation}
where $\beta$ is a primitive element in GF($2^p$). Let $n_b = np$, $k_b = kp$ and $m_b = n_b - k_b$. ${\bf{H}}_s$ has an equivalent binary image expansion ${\bf{H}}_b$, where ${\bf{H}}_b$ is an $m_b \times n_b$ binary matrix. The density of ${\bf{H}}_b$ is about 0.5 if the binary image of the parity-check matrix is directly expanded from the form given by (\ref{RS_Standard_H}). The density of the binary parity-check matrix can be reduced to about 0.3 through the sparsification method in \cite{LCC_part2}, and this relatively sparse binary parity-check matrix is denoted by ${\widetilde {\bf{H}}_b}$.

The two-dimensional product code is a simple combination of two short codes through row and column encoding. Let ${\cal C}_1$ and ${\cal C}_2$ be an $(n,k,\delta)$ RS code and a $(k_2+1,k_2,2)$ binary SPC code for the row and column encoding, respectively. The RS-SPC product codeword $\cal P$ is a rectangular array of the form:
\begin{equation}\label{RS_SPC_Product}
{\cal P} = \left[
\begin{array}{cccc}
s_{0,0} & s_{0,1} & \cdots & s_{0,n-1} \\
s_{1,0} & s_{1,1} & \cdots & s_{1,n-1} \\
\vdots & \vdots & \ddots & \vdots \\
s_{\frac{k_2}{p}-1,0} & s_{\frac{k_2}{p}-1,1} & \cdots & s_{\frac{k_2}{p}-1,n-1} \\
b_0 & b_1 & \cdots & b_{n-1}
\end{array}
 \right],
\end{equation}
where $s_{i,j}$ is the $j$-th code symbol of the $i$-th RS code, and $b_j$ is the parity bit of the $j$-th SPC code. In the rectangular array by (\ref{RS_SPC_Product}), each row is an RS codeword in ${\cal C}_1$ and each column is an SPC codeword in ${\cal C}_2$. Let the set $\left\{ \beta^0,\beta,\ldots,\beta^{p-1} \right\}$ form a basis of GF($2^p$), then $s_{i,j}$ can be expressed as $s_{i,j}=a_{i,j}^0\beta^0+a_{i,j}^1\beta+\cdots+a_{i,j}^{p-1}\beta^{p-1}$. The parity bit $b_j$ of the column $j$ can be calculated as
\begin{equation}\label{SPC_parity_bit}
{b_j} = \sum\limits_{i = 0}^{{k_2}/p - 1} {\sum\limits_{h = 0}^{p - 1} {a_{i,j}^h} },~~j \in [0:n-1],
\end{equation}
where the set $[a:b]$ is the integer subset of the set $[a,b]$.

Consider $\cal P$ consists of $L$ RS codewords. For the array by (\ref{RS_SPC_Product}), $L$ is equal to $k_2/p$ thus the rate $R_s$ of the SPC component code is $Lp/(Lp+1)$. In order to enhance the error-correcting capability, we mildly modify the structure of the $\cal P$ to lower the rate of the SPC component code. A code symbol can be split into several tuples of equal lengths $w$. Then we make each column of $\cal P$ consist of $L$ $w$-tuples (partial symbols) and a parity bit, giving the modified form as:
\begin{equation}\label{modified_RS_SPC_Product}
{\cal P} = \left[
\begin{array}{cccc}
t_{0,0} & t_{0,1} & \cdots & t_{0,\frac{pn}{w}-1} \\
t_{1,0} & t_{1,1} & \cdots & t_{1,\frac{pn}{w}-1} \\
\vdots & \vdots & \ddots & \vdots \\
t_{L-1,0} & t_{L-1,1} & \cdots & t_{L-1,\frac{pn}{w}-1} \\
b_0 & b_1 & \cdots & b_{\frac{pn}{w}-1}
\end{array}
 \right],
\end{equation}
where $t_{i,j}$ is the $j$-th $w$-tuple of the $i$-th RS code, i.e.,
\begin{equation}
t_{i,j} = \left(
{a_{i,\left\lfloor {\frac{{jw}}{p}} \right\rfloor }^{w \left( j \bmod \frac{p}{w} \right)}},
\ldots,
{a_{i,\left\lfloor {\frac{{jw}}{p}} \right\rfloor }^{w \left( j \bmod \frac{p}{w} \right)+w-1}}
\right).
\end{equation}
This modified structure gives an RS-SPC product code with rate $wLk/(wL+1)n$, denoted by ${\cal P}(n,k,w,L)$. Let ${\bf H}(n,k,w,L)$ be the parity-check matrix of ${\cal P}(n,k,w,L)$ over GF(2). Let $\Delta \left( {{{\bf{H}}},L} \right)$ be an $L \times L$ diagonal array with $L$ copies of ${\bf H}$ lying on its main diagonal and zeros elsewhere. Let $\Gamma \left( {{\bf{H}},L} \right)$ be an $1 \times L$ array with $L$ copies of ${\bf H}$. The ${\bf H}(n,k,w,L)$ can be represented as the following form:
\begin{equation}\label{H}
\begin{array}{lcl}
{\bf H}(n,k,w,L) & = & \left[
\begin{array}{c:c}
\Delta \left( {{\widetilde{\bf{H}}_b},L} \right) & {\bf O} \\ \hdashline
\Gamma \left( {{\bf{R}},L} \right) & {{\bf I}}
\end{array}\right]\\
& = & \left[
\begin{array}{ccc:c}
  \widetilde{\bf H}_b &   & &  \\
  & \ddots & &  \\
  &   & \widetilde{\bf H}_b & \\ \hdashline
  {\bf R} & \cdots & {\bf R} & {{\bf I}}
\end{array}\right],
\end{array}
\end{equation}
where $\bf I$ and $\bf O$ are the identity matrix and the all-zeros matrix with adaptive sizes, respectively, and $\bf R$ is a $\frac{pn}{w} \times np$ matrix of the following cyclic form:
\begin{equation}\label{R_matrix}
{\bf R} = \left[
\begin{array}{ccc}
  {\underbrace {1 \ldots 1}_w} &  &  \\
   & \ddots &  \\
   &  & {\underbrace {1 \ldots 1}_w}
\end{array}
\right]_{\frac{pn}{w} \times np}.
\end{equation}

\section{Iterative Soft-Decision Decoding}
In this section, we present a novel iterative soft-decision decoding scheme using the whole binary parity-check matrix given by (\ref{H}) for RS-SPC product codes, which is quite different from the well-known Chase-Pyndiah iterative decoding algorithm \cite{TPC_Pyndiah} for TPCs. In the Chase-Pyndiah algorithm, the Chase decoder is used for every component code and then the soft information can be generated from the hard-decision list. The extrinsic information in the Chase-Pyndiah algorithm is exchanged in a turbo fashion for the row and column decoders.
%On the contrast, the proposed iterative decoding scheme is based on BP, which is similar to decode an LDPC code. The BM-HDD is applied in each iteration to speed up the convergence. Since we do not use a list to reevaluate the soft information from the hard-out decoder, it is important to judge whether the decoded codeword is correct.
In the first subsection, we present a criterion for judging the codewords and speeding up the convergence. In the second subsection, a BP-based iterative decoding scheme for RS-SPC product codes is presented.

\subsection{Solution for the Undetected Errors}
Consider an RS code is transmitted on a discrete memoryless channel. Let $\varepsilon$ be the probability that a transmitted symbol is error, with an equal probability of $\varepsilon/(q-1)$ to change into one of other $(q-1)$ symbols, where $q = 2^p$. Let $P_u(E,\lambda)$ denote the probability of undetected error after correcting $\lambda$ or fewer symbol errors and it is given in \cite[Ch. 7]{LinShu}. An upper bound on the probability $P_u(E,\lambda)$ is also given in \cite[Ch. 7]{LinShu} that
\begin{equation}\label{Pu_upper_bound}
{P_u}(E,\lambda ) < {q^{ - 2t}}\sum\limits_{h = 0}^\lambda  {\dbinom{q-1}{h}} {\left( {q - 1} \right)^h}.
\end{equation}

The $P_u(E,t)$ of a high-rate RS code over a small field GF($2^p$) is relatively high, whereas an RS code over a large field GF($2^p$) usually has a quite low $P_u(E,t)$, for example, the $P_u(E,t)$ of the (255, 223, 33) RS code is upper bounded by $2.6\times10^{-14}$.

Let $\bm y = (y_0,y_1,...,y_{n_b-1})$ and ${\bm {y}}^H = (y_0^H,y_1^H,...,y_{n_b-1}^H)$ be a received vector and its hard-decision vector, respectively, where $n_b$ is the bit-length of the RS code. Let $\hat {\bm c} = ({\hat c}_0,{\hat c}_1,...,{\hat c}_{n_b-1})$ be the decoded sequence from the hard-out decoder. The soft weight $W$ of a decoded sequence is defined as a normalized ML metric of the form:
\begin{equation}\label{soft_weight}
W = {{\sum\limits_{i = 0}^{{n_b} - 1} {\left| {{y_i}} \right| \left( {y_i^H \oplus {{\hat c}_i}} \right)} }}\bigg/{{\sum\limits_{i = 0}^{{n_b} - 1} {\left| {{y_i}} \right|} }},
\end{equation}
where the symbol $\oplus$ indicates addition modulo 2. From (\ref{soft_weight}), we have $0 \le W \le 1$. Let $W_\theta$ be a soft weight threshold, where $0 \le W_\theta \le 1$. If $W < W_\theta$, the decoded sequence has a high probability to be correct. The $W_\theta$ should be carefully determined, otherwise the probability of undetected errors will be high or correct codewords will be missed.
%For the BP-based iterative decoder, the correctly determined bits can provide strong soft values to others, which will accelerate the convergence. Therefore, an appropriate $W_\theta$ is critical for the RS codes with high probability of undetected errors after the BM-HDD.

\begin{figure}[!t]
\centering
\includegraphics[width=2.5in,height=1.8in]{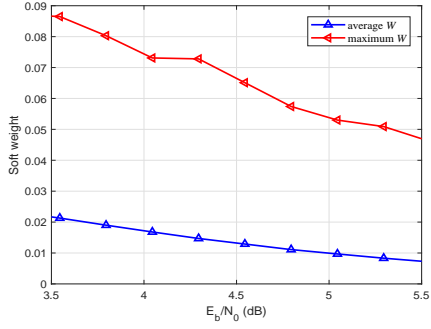}
\caption{The average and maximum soft weights of ${\cal P}(31,15,1,31)$ decoded by the genie-aided iterative decoding scheme.}
\label{SW_31_15_1_31}
\end{figure}

The threshold $W_\theta$ can be easily optimized by the simulation. For example, we use the genie-aided decoder to find the $W_\theta$ for ${\cal P}(31,15,1,31)$, where the iterative decoding algorithm as explained in later subsection is applied. The average and maximum soft weights shown in Fig. \ref{SW_31_15_1_31} are obtained by running $10^7$ decoding trials at each value of signal-to-noise ratio (SNR). The maximum soft weight is much larger than the average soft weight, and we can firstly set the soft weight threshold $W_\theta$ slightly smaller than the maximum soft weight, e.g., $W_\theta = 0.06$ for this product code. Then we can fine-tune the $W_\theta$ in the simulation, where the normal iterative decoding scheme (without genie) is applied, to achieve better performance.
For practical considerations, the cyclic redundancy check (CRC) is usually used for error detection and then the threshold-based criterion may not be necessary.

\subsection{BP-Based Iterative Decoding}
From the perspective on the LD and GD, the BM-HDD is firstly performed for each RS component codewords and then the proposed BP-based iterative decoding is applied if necessary. Assume binary phase-shift keying (BPSK) transmission over an additive white Gaussian noise (AWGN) channel with two-sided power spectral density $N_0/2$. The binary image $\bm c = (c_0, c_1,...,c_{N_{P}-1})$ of an RS-SPC product codeword is mapped into a BPSK sequence $\bm x = (x_0, x_1,...,x_{N_{P}-1})$, where $x_i=1-2c_i$, and ${N_{P}}=n_bL+n_b/w$ is the bit-length of the RS-SPC product code. Let ${\hat{\bm c}}=({\hat c}_0,{\hat c}_1,...,{\hat c}_{N_P - 1})$ be an estimation for the codeword after decoding. The received vector $\bm y = (y_0,y_1,...,y_{N_{P}-1})$ is given by
\begin{equation}\label{received_vector_AWGNC}
\bm y = \bm x + \bm w,
\end{equation}
where $\bm w$ is the noise vector with the variance ${{\sigma}_n}^2 = N_0/2$. The log-likelihood ratio (LLR) of the code bit $c_i$ is given by
\begin{equation}\label{LLR}
{\underline L}\left(c_i\right)={\rm log}\frac{{\rm Pr}\left(c_i=0|\bm y\right)}{{\rm Pr}\left(c_i=1|\bm y\right)}= \frac{2}{{\sigma_n}^2}y_i.
\end{equation}
Note that we use underlined letter $\underline L$ to distinguish from the code parameter $L$. Let $\bm{\underline{L}}=({\underline L}(c_0),{\underline L}(c_1),...,{\underline L}(c_{N_P-1}))$ denote the LLR vector. In the $\kappa_1$-th iteration, let $\underline {\bm L}^{\kappa_1} = ({\underline L}^{\kappa_1}(c_0),{\underline L}^{\kappa_1}(c_1),...,{\underline L}^{\kappa_1}(c_{N_P-1}))$ be the input LLR vector of the decoder. From the structure given by (\ref{H}), the LLR vector can be divided into $L$ sub-vectors $\underline {\bm L}_l$, $0 \le l < L$, for the RS codewords and one sub-vector $\underline {\bm L}_s$ for the parity bits.

The parity-check matrix of the RS-SPC product code consists of an upper part and a lower part. The proposed iterative decoding is a two-stage process, which utilizes the different characteristics of the upper and lower parts. We divide the check nodes into two subsets ${\cal M}^1$ and ${\cal M}^2$, where ${\cal M}^1=\{{\rm{CN-}}i,i=0,1,...,m_bL-1\}$, ${\cal M}^2=\{{\rm{CN-}}i,i=m_bL,m_bL+1,...,m_bL+n_b/w-1\}$.
Inspired by the stochastic shift iterative decoding (SSID) algorithm \cite{SSID}, we can use a shifting approach when updating the CNs $\in {\cal M}^1$. Let $\Theta \left( {\bm{\underline L},\mu} \right)$ denote the cyclically shifted version of $\bm{\underline L}$ by $\mu$ bits, where
\begin{equation}\label{cyc_shift}
\begin{array}{ll}
\Theta \left( {\bm{\underline L},\mu } \right) = & ( \underline L( {{c_{{N_P} - \mu }}} ),\underline L( {{c_{{N_P} - \mu  + 1}}} ),\ldots ,\underline L( {{c_{{N_P} - 1}}} ), \\
 & \underline L( {{c_0}} ), \ldots ,\underline L( {{c_{{N_P} - \mu  - 1}}} ) ).
\end{array}
\end{equation}
Let ${\Theta}^{-1} \left( {\bm{\underline L},\mu } \right)$ be the inverse operation of (\ref{cyc_shift}).

Let ${\cal N}(i)$ denote the set of VNs connected to the CN-$i$, and ${\cal N}(i)\backslash j$ denote the subset of ${\cal N}(i)$ without VN-$j$. Similarly, let ${\cal M}(j)$ denote the set of CNs connected to the VN-$j$, and ${\cal M}(j)\backslash i$ denote the subset of ${\cal M}(j)$ without CN-$i$.
For the CN and VN updates, we follow \cite{JN_ABP} to apply an \textit{a posteriori} probability (APP) based update scheme which reduces the decoding complexity. Let $\psi\left( \cdot \right)$ denote one iteration by the MSA to generate a sum vector of \textit{extrinsic} LLRs, denoted by $\underline {\bm L} _{ext,h}^{\kappa_1}=(\underline L_{ext,h}^{\kappa_1}(c_0),...,\underline L_{ext,h}^{\kappa_1}(c_{N_P-1}))$, for $h=1,2$, i.e.,
\begin{equation}
\underline {\bm L} _{ext,h}^{{\kappa_1}} = \mathop \psi \limits_{{\rm{CN - }}i \in {{\cal M}^h}} \left( {{{\underline {\bm L} }^{{\kappa_1}}}} \right)
\end{equation}
for $h = 1,2$. The sum of extrinsic LLRs for each bit is calculated using MSA as
%\begin{equation}\label{APP_SPA_update}
%\underline L_{ext,h}^{{k_1}}\left( {{c_j}} \right) = \sum\limits_{\updown{\scriptstyle{\rm{CN-}}i\hfill}{\scriptstyle \in {{\cal M}^h}\hfill}} {2{{\tanh }^{ - 1}}\left( {\prod\limits_{j' \in {\cal N}\left( i \right)\backslash j} {\tanh \left( {\frac{{{\underline L^{{k_1}}}\left( {{c_{j'}}} \right)}}{2}} \right)} } \right)}
%\end{equation}
%or
\begin{equation}\label{APP_MSA_update}
\begin{array}{lcl}
\underline L_{ext,h}^{{\kappa_1}}\left( {{c_j}} \right) & = & \sum\limits_{{{\rm{CN - }}i \in {{\cal M}^h}}} {\left( {\prod\limits_{j' \in {\cal N}\left( i \right)\backslash j} {{\mathop{\rm sgn}} \left( {{\underline L^{{\kappa_1}}}\left( {{c_{j'}}} \right)} \right)} } \right)}  \\
   &  & \cdot \mathop {\min }\limits_{j' \in {\cal N}\left( i \right)\backslash j} \left| {{\underline L^{{\kappa_1}}}\left( {{c_{j'}}} \right)} \right|.
\end{array}
\end{equation}

In this paper, we use the BM-HDD as the LD. Since the LD is also an initialization for the GD, we briefly present the LD process. Let $N_{err}$ denote the number of error RS component codes, and $N_{err}$ is dynamically updated in the LD and GD. Let ${\cal A}_c$ be the active set indicating the active VNs. First, set $N_{err}=L$ and ${\cal A}_c = [0:N_P-1]$. Apply the BM-HDD for each RS component code ${\cal C}_{1,l}$, for $l \in [0:L-1]$, to obtain an estimation ${\hat {\bm c}_l}$. If ${\hat {\bm c}_l} \in {\cal C}_1$ and the soft weight $W_l$ for ${\cal C}_{1,l}$ is less than the $W_\theta$, record the result and \textit{freeze} ${\cal C}_{1,l}$.

The \textit{freeze} operation means an RS component code is judged to be correctly decoded, which includes the steps as follows:
\begin{itemize}
  \item Record the decoded sequence ${{\hat {\bm c}}_l}$ of ${\cal C}_{1,l}$.
  \item Remove the positions of ${\cal C}_{1,l}$ from the active set, i.e., ${{\cal A}_c} = {{\cal A}_c}\backslash [n_bl:n_b(l+1)-1]$.
  \item Decide the LLR sub-vector $\underline {\bm L}_l$ of ${\cal C}_{1,l}$ by $\underline {\bm L}_l={\cal L}\cdot \left( {1 - 2{{\hat {\bm c}}_l}} \right)$, where $\cal L$ is a predefined positive large value.
  \item Update $N_{err} \leftarrow N_{err}-1$.
\end{itemize}

If $N_{err} > 0$ after the LD, then perform the proposed global iterative decoding. The GD can be expressed as the following steps:

\noindent{\textit{GD-1.}} Set the numbers of inner and outer iterations $\kappa_1=\kappa_2=0$, maximum numbers of inner and outer iterations $N_1$ and $N_2$, the stage-1 and stage-2 damping factors $\alpha_1$ and $\alpha_2$.

\noindent{\textit{GD-2.}} Initialize ${\underline L^0}\left( {{c_j}} \right) = \left\{ \begin{array}{l}
\frac{2}{{{{\sigma_n} ^2}}}{y_j},{\kern 1pt} {\kern 1pt} {\kern 1pt} {\kern 1pt} j \in {{\cal A}_c}\\
{\cal L} \cdot \left( {1 - 2{{\hat c}_j}} \right),{\kern 1pt} {\kern 1pt} {\kern 1pt} {\kern 1pt} j \notin {{\cal A}_c}.
\end{array} \right.$

\noindent{\textit{GD-3.}} Stage-1 CN update: Set $\mu = p\kappa_2$. Each LLR sub-vector $\underline {\bm L}_l^{\kappa_1}$ of ${\cal C}_{1,l}$ is cyclically shifted by $\mu$ bits, i.e.,
\[{\underline {\bm L}^{{\kappa_1}}} = \left( {\Theta \left( {\underline {\bm L}_0^{{\kappa_1}},\mu } \right), \ldots ,\Theta \left( {\underline {\bm L}_{L - 1}^{{\kappa_1}},\mu } \right), \underline {\bm L}_s^{{\kappa_1}}} \right).\]
Update the extrinsic LLRs, i.e.,
$\underline {\bm L}_{ext,1}^{{\kappa_1}} = \mathop \psi \limits_{{\rm{CN - }}i \in {{\cal M}^1}} \left( {{\underline {\bm L}^{{\kappa_1}}}} \right)$.
Then, shift back the extrinsic LLR sub-vector $\underline {\bm L}_{ext,1,l}^{{\kappa_1}}$ and LLR sub-vector $\underline {\bm L}_{l}^{{\kappa_1}}$ of each ${\cal C}_{1,l}$, respectively, for $0 \le l < L$, i.e.,
\[
\underline {\bm L}_{ext,1}^{{\kappa_1}} = \left( {{\Theta ^{ - 1}}\left( {\underline {\bm L}_{ext,1,0}^{{\kappa_1}},\mu } \right), \ldots ,{\Theta ^{ - 1}}\left( {\underline {\bm L}_{ext,1,L - 1}^{{\kappa_1}},\mu } \right), {\bm 0}} \right),
\]
\[
{\underline {\bm L}^{{\kappa_1}}} = \left( {\Theta^{-1} \left( {\underline {\bm L}_0^{{\kappa_1}},\mu } \right), \ldots ,\Theta^{-1} \left( {\underline {\bm L}_{L - 1}^{{\kappa_1}},\mu } \right), \underline {\bm L}_s^{{\kappa_1}}} \right).
\]

%\noindent{\textit{GD-4.}} Stage-1 VN update: Shift back the extrinsic LLR vector $\underline {\bm L}_{ext,1,l}^{{\kappa_1}}$ of each ${\cal C}_{1,l}$, for $0 \le l < L$, i.e., $\underline {\bm L}_{ext,1}^{{\kappa_1}} = \left( {{\Theta ^{ - 1}}\left( {\underline {\bm L}_{ext,1,0}^{{\kappa_1}},\mu } \right), \ldots ,{\Theta ^{ - 1}}\left( {\underline {\bm L}_{ext,1,L - 1}^{{\kappa_1}},\mu } \right)} \right)$. Then update the LLRs belonging to ${\cal A}_c$ by the extrinsic information from ${\cal M}^1$, for $0 \le j < n_bL$, ${\underline {L}^{{\kappa_1} + 1}}(c_j) = \left\{ \begin{array}{l}
%{\underline {L}^{{0}}}(c_j) + {\alpha _1}\underline {L}_{ext,1}^{{\kappa_1}}(c_j),{\kern 1pt} {\kern 1pt} {\kern 1pt} {\kern 1pt} j \in {{\cal A}_c}\\
%{\underline {L}^{{\kappa_1}}}(c_j),{\kern 1pt} {\kern 1pt} {\kern 1pt} {\kern 1pt} {\rm{otherwise}}{\rm{.}}
%\end{array} \right.$

\noindent{\textit{GD-4.}} Stage-2 CN update: $\underline {\bm L}_{ext,2}^{{\kappa_1}} = \mathop \psi \limits_{{\rm{CN - }}i \in {{\cal M}^2}} \left( {{\underline {\bm L}^{{\kappa_1}}}} \right)$.

\noindent{\textit{GD-5.}} VN update: Update the LLRs belonging to ${\cal A}_c$ by the extrinsic information from ${\cal M}^1$ and ${\cal M}^2$, for $j \le 0 < N_P$,
%\[
%{\underline {L}^{{\kappa_1} + 1}}(c_j) = \left\{ \begin{array}{l}
%{\underline {L}^{0}}(c_j) + {\alpha _1}\underline {L}_{ext,1}^{{\kappa_1}}(c_j) + {\alpha _2}\underline {L}_{ext,2}^{{\kappa_1}}(c_j), j \in {{\cal A}_c}\\
%{\underline {L}^{{\kappa_1}}}(c_j),{\kern 1pt} {\kern 1pt} {\kern 1pt} {\kern 1pt} {\rm{otherwise}}{\rm{.}}
%\end{array} \right.
%\]
\[
{\underline {L}^{{\kappa_1} + 1}}(c_j) = {\underline {L}^{0}}(c_j) + {\alpha _1}\underline {L}_{ext,1}^{{\kappa_1}}(c_j) + {\alpha _2}\underline {L}_{ext,2}^{{\kappa_1}}(c_j), j \in {{\cal A}_c}.
\]

\noindent{\textit{GD-6.}} Hard decision: $\hat {\bm c} = \frac{1}{2}\left( {1 - {\rm{sgn}}\left( {{\underline {\bm L}^{{\kappa_1} + 1}}} \right)} \right)$.

\noindent{\textit{GD-7.}} Apply BM-HDD algorithm to the active ${\cal C}_{1,l}$, for $0 \le l < L$, to obtain an estimation ${\hat {\bm c}_l}$. If ${\hat {\bm c}_l} \in {\cal C}_1$ and the soft weight $W_l$ of ${\cal C}_{1,l}$ is less than ${W_\theta }$, then \textit{freeze} ${\cal C}_{1,l}$.

\noindent{\textit{GD-8.}} If $N_{err}=0$, all the RS component codes are successfully decoded, terminate, else if $\kappa_1 = N_1-1$, go to \textit{GD-9}, otherwise set ${\kappa_1} \leftarrow {\kappa_1} + 1$ and go to \textit{GD-3}.

\noindent{\textit{GD-9.}} If $\kappa_2 = N_2-1$, terminate, otherwise set ${\kappa_2} \leftarrow {\kappa_2} + 1$, $\kappa_1 = 0$, and go to \textit{GD-2}.

Since the LLR vector of an RS code has $n$ shifted versions, $N_2$ can be set to $n$ for best performance. For convenience, we only discuss two typical values 1 and $n$ for $N_2$. Then, the iterative decoding scheme with $N_2 = 1$ or $N_2 = n$ is the low-complexity scheme or high-complexity scheme denoted by ``LCS'' or ``HCS'', respectively.

\section{Decoding Complexity}

\begin{table*}[!t]
\caption{Comparison of the Normalized Complexity}
\centering
\begin{tabular}{|c|c|c|c|c|c|}
\hline
\multirow{2}{*}{Scheme}&\multicolumn{3}{c|}{Real number computations/bit}&\multicolumn{2}{c|}{GF($2^p$) computations/bit}\\
\cline{2-6}
~&Addition&Multiplication&Comparison&BM-HDD&Parity-check\\
\hline
Pyndiah \cite{TPC_Pyndiah}& $\left( {{2^\eta } + 1.5} \right)I_{\rm avg}$ & $\left( {{2^\eta } + 1.5} \right){I_{\rm avg}}$ & $\left[ {{2^\eta } - 1 + \eta  + \frac{1}{{{n_b}}}\left( {{2^\eta } - 1 - \frac{{\left( {\eta  + 1} \right)\eta }}{2}} \right)} \right]I_{\rm avg}$ & $\frac{2^\eta}{n_b}I_{\rm avg}$ & $\frac{2^\eta(n-k)}{p}I_{\rm avg}$ \\
\hline
This paper& $(\rho m_b+1)I_{\rm avg}$ & 0 & $(\rho m_b+1)I_{\rm avg}$ & $\frac{1}{n_b} I_{\rm avg}$ & $\frac{n-k}{p}I_{\rm avg}$ \\
\hline
\end{tabular}
\label{complexity}
\end{table*}

In the following, we compare the decoding complexity for the RS-SPC product codes and TPCs \cite{TPC_Pyndiah}. We consider the TPCs with the $(n,k,\delta)$ RS component codes, denoted by ${\cal P}_R(n,k)$. We count the numbers of real number computations, including multiplications, comparisons, and additions. Each CN requires only two real number multiplications by (\ref{APP_MSA_update}) and each VN requires only one real number multiplication, which can be ignored. All modulo-2 computations are ignored. Moreover, the complexity of BM-HDD is roughly considered.

The number of real number comparisons required to update a degree-$d_c$ CN is ${d_c} + \left\lceil {{{\log }_2}{d_c}} \right\rceil  - 2$, while the number of real number additions required to update a degree-$d_v$ VN by the APP-based algorithm is $d_v$ \cite{MSA_computation}. The upper part of ${\bf H}(n,k,w,L)$ has average row weight and column weight of $\rho {n_b}$ and $\rho {m_b}$, respectively, where $\rho$ is the density of $\widetilde {\bf H}_b$. The lower part is a sparse matrix which has equal row weight and column weight of $wL+1$ and 1, respectively. The numbers of real number computations required in one iteration for the CN and VN updates are
\begin{equation}\label{CN_computations}
\begin{array}{lcl}
  N_{\rm CN} & = & {m_b}L\left( {\rho {n_b} + \left\lceil {{{\log }_2}\rho {n_b}} \right\rceil  - 2} \right)\\
  &&+ \frac{n_b}{w}\left( {wL+1 + \left\lceil {{{\log }_2}(wL+1)} \right\rceil  - 2} \right)\\
   &  \approx  & \left( {\rho {m_b} + 1} \right){n_b}L,
\end{array}
\end{equation}
\begin{equation}\label{VN_computations}
N_{\rm VN} = {n_b}L\left( {\rho {m_b} + 1} \right) + \frac{n_b}{w} \approx \left( {\rho {m_b} + 1} \right){n_b}L,
\end{equation}
respectively. In each iteration, at most $L$ soft weights are calculated which requires $\varepsilon' n_bL$ additions and $L$ multiplications. $\varepsilon'$ is the probability of the bit-error. Compared to the CN and VN updates, the number of computations required for calculating the soft weight is much smaller thus it can be ignored. Let $I_{\rm max}$ and $I_{\rm avg}$ be the maximum and average numbers of iterations, respectively, where $I_{\rm max} = N_1N_2$ for the proposed scheme. Then, the maximum number of real number computations required in the whole decoding process are approximately $2\left( {\rho {m_b} + 1} \right){n_b}LI_{\rm max}$. However, some codewords may be correctly decoded in advance and then \textit{frozen}, so we can take $2\left( {\rho {m_b} + 1} \right){n_b}LI_{\rm avg}$ to be an upper bound for the average number of real number computations. The average number of times the BM-HDD is performed for ${\cal P}(n,k,w,L)$ is upper bounded by $LI_{\rm avg}$.

Then, we consider the decoding complexity for some main procedures of the Chase-Pyndiah algorithm. Let $n_b$ be the bit-length of the RS component code in ${\cal P}_{R}(n,k)$. Suppose the Chase decoder finds $\eta$ least reliable bits and generates $2^\eta$ test patterns. Each Chase decoder performs $\left( {2{n_b} - \eta  - 1} \right)\eta /2$ comparisons to generate $2^\eta$ test patterns. For evaluating the minimum Euclidean distance from the received vector, $2^\eta n_b$ additions, $2^\eta n_b$ multiplications, and $2^\eta-1$ comparisons are required. In the procedure of calculating the extrinsic information, at most $(2^\eta-1)n_b$ comparisons are required to find the competing codeword. $2n_b$ additions and $2n_b$ multiplications are required if the competing codeword exists, otherwise $n_b$ additions and $n_b$ multiplications are required. Suppose the competing codeword can be found with a probability of 0.5, we can average the numbers of additions and multiplications to both $1.5n_b$. Therefore, the number of real number computations for one component code in one iteration is about $3 \times {2^\eta }{n_b}$. Suppose that the decoding will be terminated if all the row and column parity-checks are satisfied. Then, let $I_{\rm avg}$ be the average number of iterations, $I_{\rm avg} \ge 2$. The average number of real number computations for the Chase-Pyndiah algorithm is $3 \times {2^\eta }{n_b}nI_{\rm avg}$. In addition, the average number of times the BM-HDD is performed for TPCs is $2^\eta nI_{\rm avg}$.

Due to different component codes used in ${\cal P}(n,k,w,L)$ and ${\cal P}_R(n,k)$, we normalize the complexity to each bit. Let ${\overline N _{1,C}}$ and ${\overline N _{2,C}}$ (or ${\overline N _{1,H}}$ and ${\overline N _{2,H}}$) be the numbers of real number computations (or times the BM-HDD is performed) per bit for the proposed scheme and the Chase-Pyndiah scheme, respectively. These four normalized complexities are given as follows:
\begin{equation}\label{N_1_C}
{\overline N _{1,C}} \le 2\left( {\rho {m_b} + 1} \right)I_{\rm avg},~{\overline N _{1,H}} \le \frac{1}{n_b}I_{\rm avg},
\end{equation}
%\begin{equation}\label{N_1_H}
%{\overline N _{1,H}} \le \frac{1}{n_b}I_{\rm avg},
%\end{equation}
\begin{equation}
{\overline N _{2,C}} \approx 3 \times 2^\eta I_{\rm avg},~{\overline N _{2,H}} \approx \frac{2^\eta}{n_b}I_{\rm avg}.
\end{equation}
%\begin{equation}
%{\overline N _{2,H}} \approx \frac{2^\eta}{n_b}I_{\rm avg}.
%\end{equation}

Note that (\ref{N_1_C}) are loose upper bounds and the accurate numbers may be much smaller. ${\overline N _{1,H}}$ and ${\overline N _{2,H}}$ are just the numbers of the times the BM-HDD is performed, but the detailed complexity of the BM-HDD for different component codes deserves further consideration. We only give the detailed complexity of the BM-HDD for RS codes, which is evaluated by the number of multiplications over GF($2^p$) and it may be found in \cite{LCC_PHD}. The number of multiplications in the BM-HDD can be upper bounded by
\begin{equation}\label{Complexity_BM}
{\rm Mult}_{\rm BM-HDD} \le 7\left( n-k \right)^2 + \left( n-k \right)\left( 3n+1 \right)/2.
\end{equation}

The parity-check computations also play an important role in both two decoding schemes. $(n-k)n$ multiplications over GF($2^p$) are required for checking the codeword. For ${\cal P}(n,k,w,L)$, $(n-k)nLI_{\rm avg}$ multiplications in total or $\frac{n-k}{p}I_{\rm avg}$ multiplications per bit are required. For ${\cal P}_R(n,k)$, $2^\eta (n-k)n^2I_{\rm avg}$ multiplications in total or $\frac{2^\eta(n-k)}{p}I_{\rm avg}$ multiplications per bit are required. The above discussion for the decoding complexity is summarized in Table \ref{complexity}.

\section{Examples and Simulation Results}
%The special structure of the RS-SPC product codes preserves the low decoding latency of the LD while the error-correcting capability of the GD is comparable to codes of large blocklengths. Different from the well-known TPCs, the RS-SPC product codes offer a good trade-off between the LD and GD.
Two examples of RS-SPC product codes are given, and the simulation results and decoding complexity are discussed.
%In the following, a few applications for RS-SPC product codes with the LD and GD are briefly discussed.
%\begin{itemize}
%  \item High-speed optical communication systems: High data-rate requirement for the next-generation optical communication systems may be achieved by the two-phase decoding. These systems usually work for large SNR, thus the low-complexity LD achieves higher data-rate and lower power-consumption. The GD improves the error performance and also has a high degree of parallelism. The RS component codes give low error floors (maybe no error floors for large-distance RS component codes), which is very important for these high data-rate systems.
%  \item Massive access (uplink) and broadcasting (downlink): Suppose some relays are allowed for the uplink. These relays can decode several RS codewords from user equipments (UEs) and jointly encode them to an RS-SPC product code. The hard-decision decoding and simple SPC encoding retain low computational complexity at the relays. The error-performance improvement of the GD is valuable when the channels between the UEs and the base station are poor. The downlink is similar to a broadcasting communication, where each UE decodes a single RS codeword by the LD and performs the GD for higher reliability if necessary.
%  \item Data-storage systems: The high data-rate and low error-floor requirements are considerable for data-storage systems such as the NAND flash memory, and they also can be achieved through the two-phase decoding. The LD is important for data-storage applications with moderate-size read units.
%\end{itemize}

\begin{figure}[!t]
\centering
\includegraphics[width=2.6in,height=1.8in]{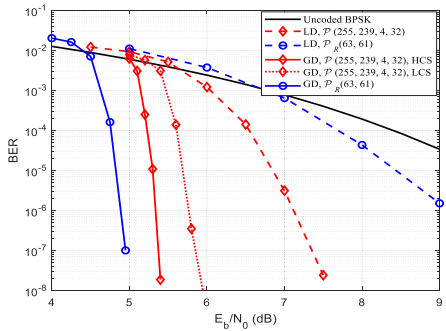}
\caption{Bit error performances of RS-SPC product code ${\cal P}(255,239,4,32)$ and the TPC ${\cal P}_R(63, 61)$.}
\label{RS_SPC_255_239}
\end{figure}

\begin{figure}[!t]
\centering
\includegraphics[width=2.6in,height=1.8in]{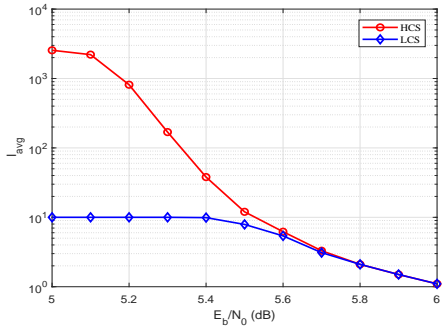}
\caption{Average number $I_{\rm avg}$ of iterations of the HCS and LCS for ${\cal P}(255,239,4,32)$.}
\label{iter_255_239}
\end{figure}

\begin{example}
\label{EXAM_255_239}
In this example, we consider the RS-SPC product code ${\cal P}(255,239,4,32)$ from the (255, 239, 17) RS code over GF($2^8$). The parameters $\alpha_1 = 0.32$, $\alpha_2 = 0.8$, $W_\theta = 0.0025$, and $N_1=10$ are set for the GD. The TPC ${\cal P}_R(63, 61)$ with 1-error-correcting RS component codes is also considered, which is one of the capacity-approaching TPCs. Chase-Pyndiah algorithm with 16 test patterns and 8 turbo iterations is applied for this TPC. The rates of ${\cal P}(255,239,4,32)$ and ${\cal P}_R(63,61)$ are 0.9300 and 0.9375, respectively. The bit-error rates (BERs) are evaluated in Fig. \ref{RS_SPC_255_239}.

At a BER of $10^{-7}$, the HCS and LCS for ${\cal P}(255,239,4,32)$ perform about 0.4 dB and 0.9 dB away from ${\cal P}_R(63, 61)$, respectively. However, the trade-off between the LD and GD is an advantage of RS-SPC product codes. It is not surprising that the LD of ${\cal P}(255,239,4,32)$ significantly outperforms the LD of ${\cal P}_R(63, 61)$ by 1.9 dB at a BER of $10^{-6}$ since the component code of this TPC can only correct 1 symbol error decoded by the BM-HDD.
%The HCS seems to have better error floors than the LCS from Fig. \ref{RS_SPC_255_239}, and no decoding errors have been found at the $E_b/N_0 = 5.5$ dB in our simulation for the HCS. Since shifting LLRs is equal to shifting the parity-check matrix, the geometry of the Tanner graph is changing for HCS thus some decoding errors caused by some specific cycles can be reduced and the error floor performance can be better. Although the LCS seems to have an error floor, we can be mildly optimistic for that the slope of the LCS curve can be reduced to at most equal to the slope of the BM-HDD curve.

Consider the decoding complexity for these two codes. the average numbers $I_{\rm avg}$ of iterations for decoding ${\cal P}(255,239,4,32)$ are shown in Fig. \ref{iter_255_239}. We can see that although the HCS performs many iterations for relatively low SNR, it only need a very small average number of iterations for large SNR. From Fig. \ref{iter_255_239}, the HCS and LCS have almost the same average complexity for large SNR. For example, at the SNR ($E_b/N_0$) of 5.8 dB, the HCS and LCS have the same $I_{\rm avg}$ of about 2. For the GD of ${\cal P}(255,239,4,32)$, the binary parity-check matrix with a density of 0.35 is used. For the TPC, we use the smallest number of iterations $I_{\rm avg} = 2$ to estimate the lowest complexity. Therefore, we can carry out the normalized complexity as follows: $\overline N _{1,C} \le 183.2$, $\overline N _{1,H} \le 9.8 \times 10^{-4}$, $\overline N _{2,C} \ge 96$, and $\overline N _{2,H} \ge 0.085$. Recall that (\ref{N_1_C}) are loose upper bounds, and we find the accurate $\overline N _{1,C}=71$ and $\overline N _{1,H} = 7.3 \times 10^{-4}$ including the BM-HDD in LD ($\overline N _{1,H} = 2.5 \times 10^{-4}$ excluding the BM-HDD in LD) for $E_b/N_0 = 5.8$ dB in the simulation. The complexity of ${\cal P}(255,239,4,32)$ are lower in both the real number computations and hard-decision decoding than the complexity of the TPC ${\cal P}_R(63, 61)$ for large SNR. We also need to explain that the complexity of the BM-HDD is only based on the number of times, but the BM-HDD for the (63, 61) RS code is much easier than decoding the (255, 239) RS code. A more accurate result may be derived from (\ref{Complexity_BM}), and the result also shows the BM-HDD of ${\cal P}(255,239,4,32)$ has lower complexity than that of ${\cal P}_R(63, 61)$.
\end{example}

\begin{figure}[!t]
\centering
\includegraphics[width=2.6in,height=1.8in]{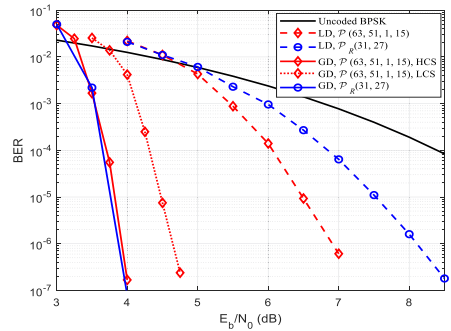}
\caption{Bit error performances of the RS-SPC product code ${\cal P}(63,51,1,15)$ and the TPC ${\cal P}_R(31,27)$.}
\label{RS_SPC_63_51}
\end{figure}

\begin{example}
In this example, we consider the RS-SPC product code ${\cal P}(63, 51, 1, 15)$ from the (63, 51, 13) RS code over GF($2^6$). ${\cal P}(63, 51, 1, 15)$ has blocklength of 6048 bits and rate 0.7589. The parameters $\alpha_1 = 0.32$, $\alpha_2 = 0.8$, $W_\theta = 0.02$, and $N_1 = 20$ are set for the GD. We also consider the TPC ${\cal P}_R(31, 27)$ from the $(31, 27, 5)$ RS code over GF($2^5$), which has comparable blocklength of 4805 and rate 0.7586. Chase-Pyndiah algorithm with 16 test patterns and 8 turbo iterations is applied for this TPC. The BERs are evaluated in Fig. \ref{RS_SPC_63_51}.

The LD of ${\cal P}(63, 51, 1, 15)$ outperforms the LD of ${\cal P}_R(31, 27)$ by 1.1 dB at a BER of $10^{-6}$. This is since different component RS codes are chosen. The HCS achieves the same performance as the Chase-Pyndiah algorithm for ${\cal P}_R(31, 27)$, while the LCS performs about 0.7 dB from ${\cal P}_R(31, 27)$. The HCS for this RS-SPC product code also has low average complexity for large SNR, e.g., it takes an average of 2 iterations to converge at $E_b/N_0 = 4.5$ dB.

%In addition, if the system employs CRC, the threshold-based criterion may not be needed. For example, we find 16-bit CRC is effective to avoid undetected errors for this code in the simulation, thus $W_\theta$ can be set to 1.
\end{example}

\section{Conclusion}
In this paper, we present a BP-based iterative decoding scheme for RS-SPC product codes. These codes can be easily constructed from various RS codes, while TPCs are difficult to employ many widely-used RS codes as the component codes such as the RS codes of length 255. The only problem for RS-SPC product codes is to control the probability of undetected errors for the hard-decision decoding. Better component codes can be used for RS-SPC product codes, which gives a good performance for LD. The two-phase decoding scheme preserves the low decoding latency of the LD while the error-correcting capability of the GD is comparable to codes of large blocklengths. This flexible structure may meet the high-reliability and low-latency requirement for future communication systems.

%There are several aspects of RS-SPC product codes which may deserve further investigation. The hard-decision decoding is applied for each iteration, but it is difficult to output soft information. We only utilize the hard decisions of reliable component codewords and freeze them in the subsequent iterations. This operation incurs undetected errors for the RS codes with small minimum distances. Pyndiah's soft decision algorithm \cite{TPC_Pyndiah} is an effective method for the list decoding, and it may deserve further research for RS-SPC product codes. Another interesting aspect is the code construction. Larger-dimension RS-SPC product codes have larger minimum distances, and they may be potential to improve the error performance particularly when short RS component codes are used.

\appendices

%\section{}

\section*{Acknowledgment}
This work was supported by the National Natural Science Foundation of China under Grant 61771133 and the Jiangsu Province Basic Research Project under Grant BK20192002.

\ifCLASSOPTIONcaptionsoff
  \newpage
\fi

\end{document}